\documentclass[osajnl,twocolumn,showpacs,superscriptaddress,10pt]{revtex4-1} 
\usepackage{amsmath,amssymb,graphicx,subfigure}

\usepackage{color}

\begin{document}

\title{Absolute Frequency Measurement of Rubidium 5S-7S Two-Photon Transitions}

\author{Piotr Morzy\'nski}
\affiliation{Institute of Physics, Faculty of Physics, Astronomy and Informatics, Nicolaus Copernicus University, Grudzi\c{a}dzka 5, PL-87-100, Toru\'n, Poland}
\author{Piotr Wcis\l{}o}
\affiliation{Institute of Physics, Faculty of Physics, Astronomy and Informatics, Nicolaus Copernicus University, Grudzi\c{a}dzka 5, PL-87-100, Toru\'n, Poland}
\author{Piotr Ablewski}
\affiliation{Institute of Physics, Faculty of Physics, Astronomy and Informatics, Nicolaus Copernicus University, Grudzi\c{a}dzka 5, PL-87-100, Toru\'n, Poland}
\author{Rafa\l{} Gartman}
\affiliation{Institute of Physics, Faculty of Physics, Astronomy and Informatics, Nicolaus Copernicus University, Grudzi\c{a}dzka 5, PL-87-100, Toru\'n, Poland}
\author{Wojciech Gawlik}
\affiliation{Institute of Physics, Faculty of Physics, Astronomy and Informatics, Jagiellonian University, Reymonta 4, PL-30-059 Krak\'ow, Poland}
\author{Piotr Mas\l{}owski}
\affiliation{Institute of Physics, Faculty of Physics, Astronomy and Informatics, Nicolaus Copernicus University, Grudzi\c{a}dzka 5, PL-87-100, Toru\'n, Poland}
\author{Bart\l{}omiej Nag\'orny}
\affiliation{Institute of Physics, Faculty of Physics, Astronomy and Informatics, Nicolaus Copernicus University, Grudzi\c{a}dzka 5, PL-87-100, Toru\'n, Poland}
\author{Filip Ozimek}
\affiliation{Institute of Experimental Physics, Faculty of Physics, University of Warsaw, Ho\.za 69, PL-00-681 Warsaw, Poland}
\author{Czes\l{}aw Radzewicz}
\affiliation{Institute of Experimental Physics, Faculty of Physics, University of Warsaw, Ho\.za 69, PL-00-681 Warsaw, Poland}
\author{Marcin Witkowski}
\affiliation{Institute of Physics, Faculty of Physics, Astronomy and Informatics, Nicolaus Copernicus University, Grudzi\c{a}dzka 5, PL-87-100, Toru\'n, Poland}
\affiliation{Institute of Physics, University of Opole, Oleska 48, PL-45-052 Opole, Poland}
\author{Roman Ciury\l{}o}
\affiliation{Institute of Physics, Faculty of Physics, Astronomy and Informatics, Nicolaus Copernicus University, Grudzi\c{a}dzka 5, PL-87-100, Toru\'n, Poland}
\author{Micha\l{} Zawada} \email{Corresponding author: zawada@fizyka.umk.pl}
\affiliation{Institute of Physics, Faculty of Physics, Astronomy and Informatics, Nicolaus Copernicus University, Grudzi\c{a}dzka 5, PL-87-100, Toru\'n, Poland}

\begin{abstract}
We report the absolute frequency measurements of rubidium 5S-7S two-photon transitions with  a cw laser digitally locked to an atomic transition and referenced to an optical frequency comb. The narrow, two-photon transition, 5S-7S (760 nm)  insensitive to first order in a magnetic field, is a promising candidate for frequency reference. The performed tests yield the transition frequency with accuracy 
better than reported previously. 
\end{abstract}


\maketitle 

\section{Introduction}
The International Committee for Weights and Measures recommended several radiations for the practical realization of the metre. At the border of the visible and near-infrared ranges, in particular,  the International Bureau of Weights and Measures (BIPM) recommends the 5S${}_{1/2}$(F=3) - 5D${}_{5/2}$(F=5) two-photon transition in ${}^{85}$Rb with a standard uncertainty of 5 kHz (the relative standard uncertainty of $1.3 \times 10^{-11}$)~\cite{BIPM}. Recent development in phase-stabilized optical frequency combs based on mode-locked femtosecond lasers allowed determination of the absolute frequency of a similar transition in Rb, 5S${}_{1/2}$-7S${}_{1/2}$, which is 100 times weaker  than 5S-5D, yet less sensitive to stray magnetic fields. 
At the 5S-5D transition the rubidium atoms must be carefully shielded against the magnetic field to avoid any linear Zeeman shifts. On the other hand, the 5S and 7S levels have the same Land\'e $g$ factors which cancels the linear Zeeman shifts in the 5S-7S transition. The ac-Stark effect in the 5S-7S transition is also smaller than in the 5S-5D case. 
 Because of the difference in the signal strengths, all previous measurements of the 5S-7S transition~\cite{Ko04,Marian05,Pandey08,Barmes13} yielded absolute values of its frequency less accurate than for the 5S-5D transition~\cite{Touahri97}.
 In this work we report the measurement of the absolute frequency measurements of the 5S${}_{1/2}$-7S${}_{1/2}$ transitions in ${}^{87}$Rb and ${}^{85}$Rb with relative standard uncertainty smaller than $10^{-11}$ which is 
better than measured previously and is comparable with the measurements of the 5S-5D transition~\cite{Touahri97}.

\section{Experimental arrangement}

\begin{figure}[!t]
\centering
\includegraphics[width=0.9\columnwidth]{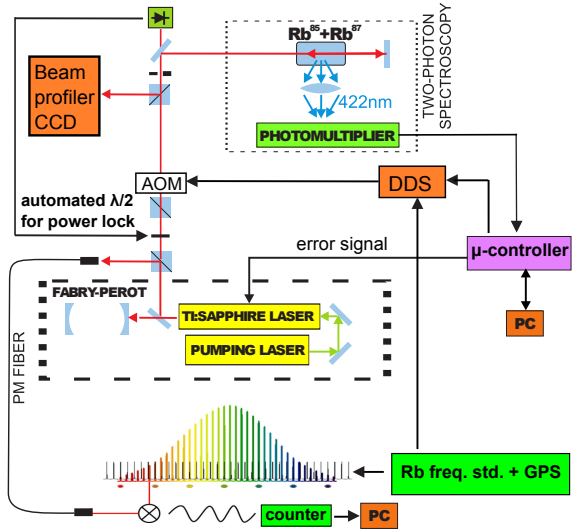}
\caption{Experimental setup.}
\label{fig:setup}
\end{figure}

The experimental setup is shown in Fig. \ref{fig:setup}.
We have used a commercial ring-cavity titanium
sapphire (TiSa) laser to study the two-photon 5S${}_{1/2}$-7S${}_{1/2}$ transitions at 760 nm in a hot rubidium vapour cell at temperatures up to $140^{\circ}$C. The TiSa laser, pre-stabilised by a Fabry-P\'erot cavity, has the linewidth of 300 kHz. The two-photon spectroscopy signal is observed by a photomultiplier tube in the 7S-6P-5S radiative cascade, with the 6P-5S blue fluorescence around 421 nm.  The laser is digitally locked to the measured transition. The digital lock, a technique generally used in optical atomic clocks, allows for long interrogation of the non-modulated reference beam with an optical frequency comb. That technique and good long-term stability of our system reduce the statistical uncertainties of measured frequencies below 1~kHz, an order of magnitude better than in Refs.~\cite{Ko04,Marian05,Pandey08}, thanks to the longer averaging times. The low statistical uncertainty enabled studies of systematic shifts with accordingly higher precision.

\begin{figure}[!t]
\centering
\includegraphics[width=0.9\columnwidth]{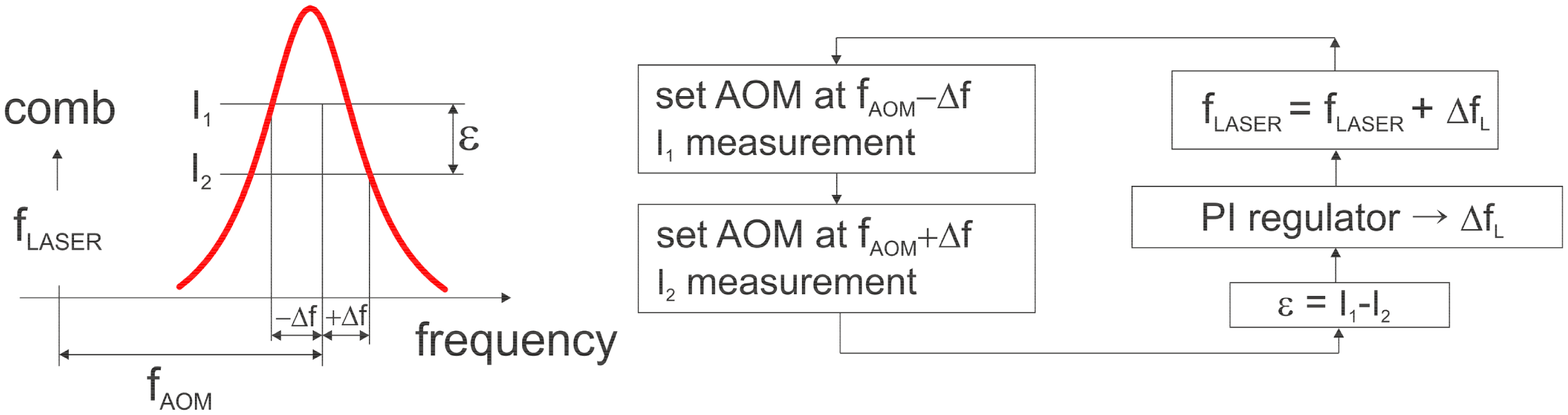}
\caption{The scheme of the digital lock.}
\label{fig:digilock}
\end{figure}

The idea of the digital lock is depicted in Fig. \ref{fig:digilock}. An acousto-optic modulator (AOM) driven by a direct digital synthesizer (DDS) square-wave modulates the light frequency with the step $2\Delta f$ equal to the half-width of the line. 
The AOM carrier frequency $f_{AOM}$ is chosen such that the AOM efficiency with $f_{+}=f_{AOM}+\Delta f$ and $f_{-}= f_{AOM}-\Delta f$ is the same.
The microcontroller (Atmel AT91SAM7S), which controls the DDS, counts the photomultiplier pulses from the fluorescence signal of the 6P-5S transition. The error signal for the laser lock is calculated from the difference of the counts corresponding to fluorescence for $f_{+}$ and $f_{-}$. The software PI regulator in the microcontroller calculates the correction $\Delta f_L$ and applies it to the TiSa laser.
 Switching between the $f_{\pm}$ frequencies in the DDS is completed in 150 ns and the acoustic wave needs few $\mu$s to completely propagate the frequency changes through the AOM PbMo0${}_4$ crystal. The following acquisition of the photomultiplier pulses takes 38 ms which is long enough to ignore chirping effects from switching the $f_{\pm}$ frequencies.

The power of the light sent to the rubidium cell is stabilised by the software PI regulator on the embedded PC (FOX Board G20) with a half-wave plate mounted on a piezo-driven mount and a polariser. To exclude the residual Doppler  effect, caused by the wave-front curvature,   the counter-propagating beams in the two-photon spectroscopy are not focused and their relative positions are controlled by a CCD beam profiler. 

\begin{figure}[!t]
\centering
\includegraphics[width=0.7\columnwidth]{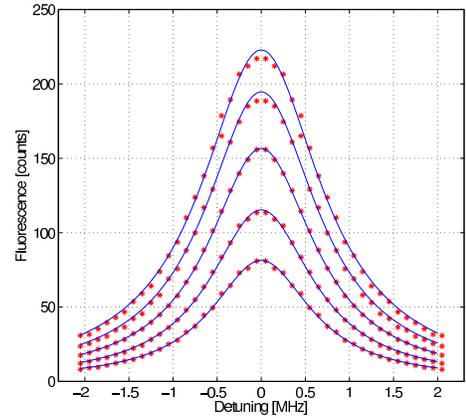}
\caption{The ${}^{87}$Rb F=2-F'=2 line profile for different intensities of the probing light. The Lorentz profile is fitted to the measured data.}
\label{fig:line}
\end{figure}

 The digital lock can be also used to measure the line profile (Fig. \ref{fig:line}) by changing $\Delta f$ in the digital lock algorithm and recording accordingly the fluorescence intensity. Since the counter-propagating beams are not focused, the transition is characterized  by  a power-broadened Lorentzian profile. The measurement accuracy may be reduced at the line center
where the digital lock is less precise because of small value of the fluorescence derivative with respect to the laser frequency
 leading to underestimation of measured fluorescence at given frequency.

\begin{figure}[!t]
\centering
\includegraphics[width=0.9\columnwidth]{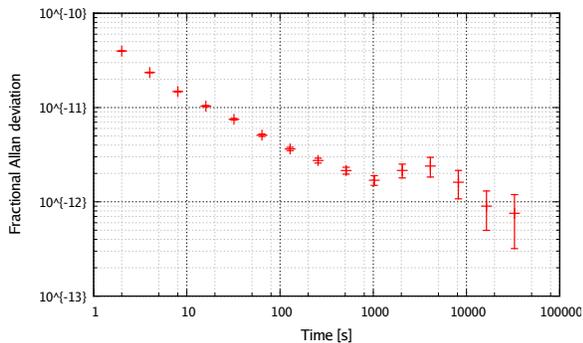}
\caption{Allan variance of the frequency of the TiSa laser locked to the ${}^{87}$Rb F=2-F'=2 transition measured with the optical frequency comb.}
\label{fig:allan}
\end{figure}

Part of the TiSa light is sent directly to the Er-dopped fiber optical frequency comb (Menlo FC1500-250-WG). The comb, the DDS synthesizers and counters in the experiment are locked to a microwave Rb frequency standard (SRS FS725), disciplined by the GPS (Connor Winfield FTS 375) with fractional freuqency uncertainty better than $1\times 10^{-12}$ at our measurement times.
The fractional Allan variance of the frequency of the TiSa laser locked to the ${}^{87}$Rb F=2-F'=2 transition measured with the optical frequency comb is presented in Fig. \ref{fig:allan}. After 1000 s the Rb frequency standard reaches its final stability, and our system is further disciplined by the GPS which improves the stability, as seen in the plot, after $10^{4}$ s integration time.

\section{Results}

Several systematic effects should be
taken into account to deduce the transition frequency. Most pronounced is the systematic pressure shift of the measured transition. 

\begin{figure}[!t]
\centering
\includegraphics[width=0.8\columnwidth]{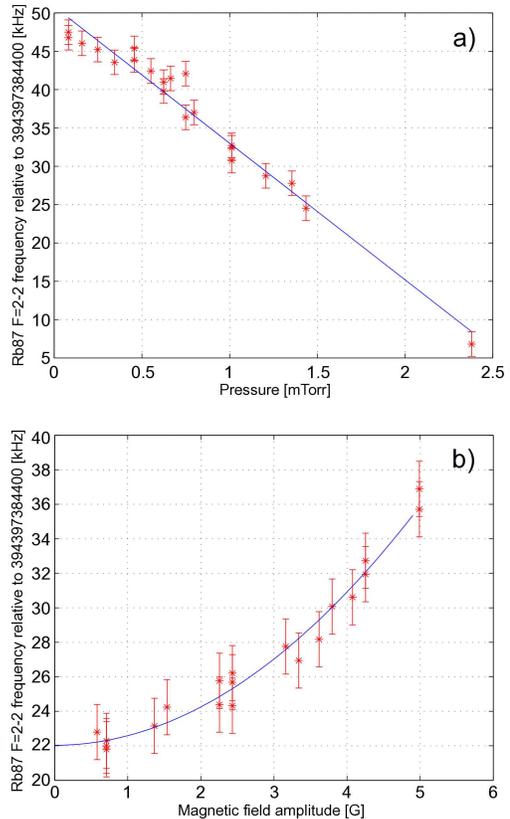}
\caption{Calibration of  a) the pressure shift b) quadratic Zeeman shift 	of the ${}^{87}$Rb F=2-F'=2 transition.}
\label{fig:press}
\end{figure}

The rubidium vapour pressure is determined by
the cell temperature~\cite{Alcock}. We measured the absolute frequencies at different temperatures (accuracy of 0.2~K)  and estimated the 
coefficient of the pressure shift interaction Rb-Rb as
-17.82(81) kHz/mTorr (Fig. \ref{fig:press}a).
  This value differs from the value in Ref.~\cite{Ko04} since we used a more accurate expression~\cite{Alcock} for calculating the vapour pressure at given temperature.
Finally, the shift was extrapolated to the zero pressure.
  In the same way we measured and  calibrated the ac-Stark shift. 
Surprisingly, the measured ac-Stark shift is much lower than calculated in Ref.~\cite{Ko04}. Analysis of this effect is, however, beyond the scope of this report.
 In addition to the Rb-Rb pressure
shift, additional contribution comes from collisions with residual gas present in the cell (mainly Ar, since rubidium acts as a getter for other impurities).
In recent work of Wu et al.~\cite{Wu13} the 6S-8S transition frequency measured in 10 different commercial Cs vapour cells varied by hundreds of kHz, but measurements of the 5S-5D transition~\cite{Touahri97} proved that in Rb this effect is not that big.
 Our Rb cell has been filled under vacuum of the order of $10^{-8}$ Torr but we assume it could drop to about $10^{-4}$ Torr  because of the cell ageing.
The Rb-Ar collisional shift was estimated assuming van der Waals  interaction~\cite{Trawinski06} and verified with data in Ref. \cite{Weber82}.
 Fig. \ref{fig:press}b depicts the determination of the quadratic Zeeman shift coefficient measured by application of an external, calibrated magnetic field.  The actual magnetic field was measured by a precise magnetometer.
  The black body radiation \cite{Farley} and second order Doppler shifts were calculated for a given stabilised cell temperature.	

 The systematic shift caused by different efficiencies of the AOM probing two sides of the transition is most likely the fundamental limit of the measured uncertainty in the digital lock scheme. This shift
is dependent on the laser frequency derivatives of fluorescence corresponding to $f_{+}$ and $f_{-}$.
 This effect is most clearly seen when the residual Doppler effect caused by some misalignment of the probe beams broadens the line by few MHz. The resulting change of
these derivatives shifts the transition by tens of kHz. This shift is quantified by varying the $\Delta f$ value in the DDS lock and the AOM diffraction order (+$f_{AOM}$ or -$f_{AOM}$). These systematic shifts can be significantly reduced in future measurements. For example, in the optical clocks this effect is eliminated by the injection locked slave laser.

The accuracy budget for typical experimental conditions, i.e. laser intensity of 11~W/cm${}^2$ in each beam,  beams $1/e^2$ diameter of 1.4~mm and temperature of $128.5^{\circ}$C, is presented in Table \ref{tab:acc}.

\begingroup
\squeezetable
\begin{table}[h!]

\caption{Accuracy budget for typical experimental   conditions
(m - measured, c - calculated).}
\label{tab:acc}
\centering

\begin{tabular}{ccc}
            \hline 
            Effect & Shift [kHz] & Uncert. [kHz] \\ 
            \hline \hline
            Pressure shift due to Rb-Rb interaction${}^{m}$ & -24.14 & 1.1 \\ 
            \hline 
            Light shift${}^{m}$& 0.4 & 1.6 \\ 
            \hline 
            Quadratic Zeeman Shift${}^{m}$& 0.055& 0.034\\
            \hline
            Line pulling${}^{c}$& 0 & 0.01 \\ 
            \hline 
             Pressure shift due to background gases${}^{c}$& -0.75 & 0.75 \\ 
            \hline 
            Second order Doppler effect${}^{c}$& -0.168 & 0.001 \\ 
            \hline 
             DDS electronics \&  digital lock${}^{m}$& -6.40 &  1.2 \\ 
            \hline 
            Black Body Radiation${}^{c}$& -0.666 & 0.004 \\ 
            \hline 
            Rb frequency standard \& GPS${}^{c}$& 0 & 0.4 \\ 
            \hline \hline 
             Total: &  -31.7 &  2.4\\
            \hline
            \end{tabular} 
\end{table}
\endgroup

\begin{table}[h!]
\caption{Measured absolute frequencies of the two-photon 5S${}_{1/2}$-7S${}_{1/2}$ transitions.}
\label{tab:res}
\centering

\begin{tabular}{cc}
            \hline 
            Transitions  & Frequency [kHz] \\ 
            \hline            
            ${}^{85}$Rb F=2-F'=2 ~~ & 394399282854.2(2.4)\\
            ${}^{85}$Rb F=3-F'=3 ~~ & 394397907005.6(2.8)\\
            ${}^{87}$Rb F=1-F'=1 ~~ & 394400482036.5(3.8)\\
            ${}^{87}$Rb F=2-F'=2 ~~ & 394397384443.1(2.6) \\
            \hline 
          \end{tabular}
          
\end{table}

\begin{figure}[!t]
\centering
\includegraphics[width=0.8\columnwidth]{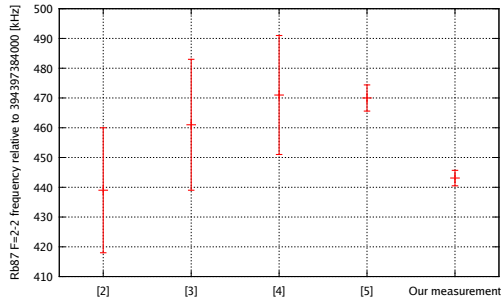}
\caption{Comparison of the ${}^{87}$Rb F=2-F'=2 transition frequency with the previously known values.}
\label{fig:zest}
\end{figure}

\begin{table}[h!]

\caption{Hyperfine A constants of the Rb 7S${}_{1/2}$ state and isotope shift of the 5S${}_{1/2}$-7S${}_{1/2}$ transition.}
\label{tab:hyp}
\centering

\begin{tabular}{ccc}            
\hline            
            Hyperfine A constants  [kHz]& & \\
            \hline 
            ${}^{85}$Rb 7S & 94678.4(2.3) & our measurement\\
                           & 94680.7(3.7) & \cite{Barmes13}\\
                           & 94658(19) & \cite{Ko04}\\                       
            ${}^{87}$Rb 7S & 319747.9(2.3)& our measurement\\
                           & 319751.8(5.1) & \cite{Barmes13}\\
                           & 319759(28) & \cite{Ko04}\\
                           & 319702(65) & \cite{Marian05}\\
            \hline 
            5S-7S Isotope shift  [kHz]& & \\
            \hline 
            ${}^{85}$Rb - ${}^{87}$Rb &131529.6(6.6)& our measurement\\
                                      &131533(15) & \cite{Barmes13}\\
                                      &131567(73) & \cite{Ko04}\\
            \hline 
            \end{tabular} 
\end{table}

In Table \ref{tab:res} we present the measured frequencies of four 5S${}_{1/2}$-7S${}_{1/2}$ transitions in ${}^{87}$Rb and ${}^{85}$Rb.
Comparison of the measured frequency  of the ${}^{87}$Rb F=2-F'=2 transition with the previously measured values is depicted in Fig. \ref{fig:zest}. With the knowledge of the hyperfine splitting of the 5S${}_{1/2}$ state~\cite{Bize}
the hyperfine A constants of the 7S${}_{1/2}$ state were derived, as well as the 5S${}_{1/2}$-7S${}_{1/2}$ transition isotope shift.  The results are presented in Table \ref{tab:hyp}.
Comparisons presented in Fig. \ref{fig:zest} and Table \ref{tab:hyp} show  that while there is a good agreement of all previous measurements of the A hyperfine constants and isotope shift of the 5S-7S transition, the absolute frequency measurements agree within the expanded uncertainties only with the values given in Refs.~\cite{Ko04,Marian05,Pandey08}
and  all transition frequencies measurements in Ref.~\cite{Barmes13} are systematically higher that those measured in Ref.~\cite{Ko04}, where the measurement
scheme was similar to ours.
Since the measurements of the hyperfine and isotope structures are not
absolute but relative, the discrepancy between Ref.~\cite{Barmes13} and the present work seems to indicate a systematic shift of the
5S-7S transition that remains to be identified.

\section{Conclusion}

We have performed a series of measurements of the absolute frequency of the 5S${}_{1/2}$-7S${}_{1/2}$ two-photon transitions in rubidium vapour with an optical frequency comb. We have also estimated the  A constants of the hyperfine splitting of the 7S${}_{1/2}$ state and the isotope shift between ${}^{85}$Rb and ${}^{87}$Rb of the 5S${}_{1/2}$-7S${}_{1/2}$ transition. 
Thanks to  low statistical uncertainty and thorough study of all systematic shifts, 
the accuracy of the 5S${}_{1/2}$-7S${}_{1/2}$ transition frequency obtained in the present work is higher than previously reported~\cite{Ko04,Marian05,Pandey08,Barmes13}.

\section*{Acknowledgment}

This work has been performed in the National Laboratory FAMO in Toru\'{n} and supported by
 the Polish National Science Centre Project No. 2012/07/B/ST2/00235  
and the TEAM Projects of the FNP, co-financed by the EU within the  European Regional Development Fund.

\eject

\end{document}